\documentstyle[12pt]{article}
\def\1ad{\mbox{\normalsize $^1$}}
\def\2ad{\mbox{\normalsize $^2$}}
\def\3ad{\mbox{\normalsize $^3$}}
\def\4ad{\mbox{\normalsize $^4$}}
\def\5ad{\mbox{\normalsize $^5$}}
\def\6ad{\mbox{\normalsize $^6$}}
\def\7ad{\mbox{\normalsize $^7$}}
\def\8ad{\mbox{\normalsize $^8$}}
\def\makefront{\vspace*{1cm}\begin{center}
\def\newtitleline{\\ \vskip 5pt}
{\Large\bf\titleline}\\
\vskip 1truecm
{\large\bf\authors}\\
\vskip 5truemm
\addresses
\end{center}
\vskip 1truecm
{\bf Abstract:}
\abstracttext
\vskip 1truecm}
\def\a{\alpha}
\def\b{\beta}
\def\g{\gamma}
\def\e{\epsilon}
\def\f{\varphi}
\def\F{\Phi}
\def\k{\kappa}
\def\dt#1{{\buildrel {\hbox{\LARGE .}} \over {#1}}}     
\def\ad{\dt{\a}}
\def\da{\dt{\a}}
\def\bd{\dt{\b}}
\def\gd{\dt{\g}}
\def\kd{\dt{\k}}

\def\cd{{\cal D}}
\def\cp{{\cal P}}
\DeclareFontFamily{OT1}{msb}{}{}
\DeclareFontShape{OT1}{msb}{m}{n}
 {  <5> <6> <7> <8> <9> <10> gen * msbm
      <10.95><12><14.4><17.28><20.74><24.88>msbm10}{}
\DeclareMathAlphabet{\bubble}{OT1}{msb}{m}{n}
\def\bR{{\bubble R}}
\def\bZ{{\bubble Z}}

\def\pa{\partial}

\def\fr#1#2{{\textstyle{#1\over\vphantom2\smash{\raise.20ex
        \hbox{$\scriptstyle{#2}$}}}}}                   
\def\bo{{\raise-.5ex\hbox{\large$\Box$}}}
\def\Tr{{\rm Tr}}
\def\ket#1{\left| #1\right\rangle}

\begin {document}
hep-th/9801117 \hfill ITP-UH-01/98
\def\titleline{
On the self-dual geometry of N=2 strings
}
\def\authors{
Chandrashekar Devchand \1ad , Olaf Lechtenfeld \2ad
}
\def\addresses{
\1ad
Max-Planck-Institut f\"ur Mathematik in den Naturwissenschaften\\
Inselstra\ss{}e 22-26, D-04103 Leipzig\\
\2ad
Institut f\"ur Theoretische Physik, Universit\"at Hannover\\
Appelstra\ss{}e 2, D-30167 Hannover\\
}
\def\abstracttext{
We discuss the precise relation of the open $N{=}2$ string to 
a self-dual Yang-Mills (SDYM) system in $2{+}2$ dimensions.
In particular, we review the description of the string target space action 
in terms of SDYM in a ``picture hyperspace'' parametrised by the standard
vectorial $\bR^{2,2}$ coordinate together with  a commuting spinor of 
$SO(2,2)$. The component form contains an infinite tower of prepotentials 
coupled to the one representing the SDYM degree of freedom. The truncation 
to five fields yields a novel one-loop exact lagrangean field theory.
}
\makefront
\noindent{\bf 1. Introduction\ \ } The relation to self-dual Yang-Mills (SDYM) 
of the critical open $N{=}2$ string has recently been elaborated by 
us~\cite{dl} in view of the particular picture degeneracy and global $SO(2,2)$ 
properties of the physical spectrum of string states. There has been much
discussion in the literature of this relationship since Ooguri and 
Vafa~\cite{ov} first mooted the idea that the self-duality equations, 
\begin{equation}
F_{\mu\nu} - \fr12\e_{\mu\nu}^{\;\;\;\;\rho\lambda} F_{\rho\lambda}\ =\ 0\ ,
\label{1}
\end{equation}
with the field strengths taking values in the Chan-Paton Lie algebra,
describe what at that stage appeared to be the single dynamical
degree of freedom of the open $N{=}2$ string. (We shall not give all
relevant references here, referring to~\cite{dl} for further references).
The comparison has been based on determinations of tree-level amplitudes for 
the two theories, so light-cone gauge action principles for SDYM have played a 
central role in the discussion. In two-spinor notation, using the splitting of 
the $\bR^{2,2}$ ``Lorentz algebra'',
\begin{equation}
so(2,2)\ \cong\ sl(2,\bR) \oplus sl(2,\bR)' 
\quad\Longleftrightarrow\quad
x^\mu \sigma_\mu^{\a\bd}\ =\ x^{\a\bd}\ =\ 
{\textstyle\left(\begin{array}{cc}
x^0{+}x^3 & x^1{+}x^2 \\ x^1{-}x^2 & x^0{-}x^3
\end{array}\right)}\ ,
\label{2}
\end{equation}
the three (real) SDYM equations take the form
\begin{equation}
F_{\a\b}\ \equiv\ \fr12 \left(\pa_{(\a}^{\;\;\;\gd} A_{\b)\gd}\ +\
[ A_\a^{\;\;\gd} , A_{\b\gd} ] \right) \ =\ 0 
\label{3}
\end{equation}
In components, with the spinor indices $\a,\b$ taking values $+$ and $-$,
we have
\begin{eqnarray}
 F^{++} &\equiv& \pa^{+\gd}A^+_{\;\;\gd} 
+ \fr12 [A^{+\gd},A^+_{\;\;\gd}] \ = 0 \nonumber \\
 F^{+-} &\equiv& \fr12 \left( \pa^{+\gd}A^-_{\;\;\gd}
+ \pa^{-\gd}A^+_{\;\;\gd} 
+ [A^{+\gd},A^-_{\;\;\gd}] \right)\ = 0 \\
 F^{--} &\equiv& \pa^{-\gd}A^-_{\;\;\gd} 
+ \fr12 [A^{-\gd},A^-_{\;\;\gd}] \ = 0 \quad. \nonumber
\label{4}
\end{eqnarray}
Clearly, the $(++)$ equation affords the generalised light-cone gauge
$A^+_{\gd}{=}0$ in which $F^{+-}$ becomes homogeneous.
Two strategies now suggest themselves.
First, resolving the (inhomogeneous) $(--)$ equation in the Yang fashion,
\begin{equation}
A^-_{\;\;\ad}\ =\ e^{-\phi}\ \pa^-_{\;\;\ad}\ e^{+\phi}\ ,
\label{5}
\end{equation}
the $(+-)$ equation describes the $\phi$-dynamics in the form of the 
(non-polynomial) Yang equation
\begin{equation}
\pa^{+\ad}\ (e^{-\phi}\ \pa^-_{\;\;\ad}\ e^{+\phi})\ =\ 0\ .
\label{6}
\end{equation}
Second, the (homogeneous) $(+-)$ equation is instead fulfilled in terms of
a Leznov prepotential, writing
\begin{equation}
A^-_{\;\;\ad}\ =\ \pa^+_{\;\;\ad}\ \f^{--}\ ,
\label{7}
\end{equation}
which then must satisfy $F^{--}=0$, tantamount to the (quadratic) Leznov 
equation, 
\begin{equation}
\bo \f^{--} - \fr12 [\pa^{+\ad} \f^{--} , \pa^+_{\;\;\ad} \f^{--}]\ =\ 0\  .
\label{8}
\end{equation}
The light-cone gauge explicitly breaks the global $SO(2,2)$ covariance
of eq. (\ref{3}) to $GL(1,\bR)\otimes SL(2,\bR)'$.
In a Cartan-Weyl basis for $sl(2,\bR)$ consisting of a diagonal hyperbolic
generator $L_{+-}$ and two parabolic generators $L_{\pm\pm}$, the unbroken
$gl(1,\bR)$ generator is $L_{+-}$ in the Yang but $L_{++}$ in the Leznov case.

Non-covariant action principles for (\ref{6}) or
(\ref{8}) yield themselves using merely the prepotentials, 
\begin{eqnarray}
S_{\rm Yang} &=& \mu^2 \int\!d^4x\; \Tr\biggl\{ -\fr12\ \phi\bo\phi\ 
+\ \fr13\ \phi\ \pa^{(+\ad}\phi\ \pa^{-)}_{\;\;\ad} \phi\ 
+\ {\cal{O}}(\phi^4) \biggr\} 
\label{9} \\
S_{\rm Leznov} &=& \mu^2 \int\!d^4x\; \Tr\biggl\{ -\fr12\ \f^{--}\bo\f^{--}\ 
+\ \fr16\ \f^{--}\Bigl[ \pa^{+\ad}\f^{--},\pa^+_{\;\;\ad}\f^{--}\Bigr]\biggr\}
\label{10}
\end{eqnarray}
with some mass scale $\mu$.
Alternatively, Lagrange multipliers facilitate the construction of
dimensionless actions, for example,
\begin{equation}
S_{\rm CS}\ =\ \int\!d^4x\; \Tr\biggl\{ -\f^{++}\bo\f^{--}\
+\ \f^{++}\Bigl[ \pa^{+\ad}\f^{--} , \pa^+_{\;\;\ad}\f^{--}\Bigr]\biggr\}
\label{11}
\end{equation}
which was shown to be even one-loop exact by Chalmers and Siegel~\cite{cs}.

The tree-level amplitudes following from these actions are extremely simple. 
Since we are dealing with massless fields in $2{+}2$ dimensions, the
on-shell momenta factorise, 
\begin{equation}
k^{\a\bd}\ k_{\a\bd}\ =\ 0 \qquad\qquad\Longleftrightarrow\qquad\qquad
k^{\a\bd}\ =\ \k^{\a}\ \k^{\ad} \quad.
\label{13}
\end{equation}
The on-shell three-point functions $A_3(k_1,k_2,k_3)$ can be read off as
\begin{equation}
A_3^{\rm Yang}\ =\ f^{abc}\ \k_1^{(+}\k_2^{-)}\ \k_1^\ad \k_{2\ad} 
\qquad,\qquad
A_3^{\rm Leznov}\ =\ f^{abc}\ \k_1^+\k_2^+\ \k_1^\ad \k_{2\ad}
\label{14}
\end{equation}
where $\sum_i k_i=0$ and $f^{abc}$ are the structure constants of the gauge
group. Surprisingly, the four-point Feynman diagrams sum to zero, in virtue 
of a quartic contact interaction in the Yang case. It is believed that all 
higher tree amplitudes vanish on-shell. The version of Chalmers and Siegel 
leads to the same tree-level amplitudes as the Leznov action, although in the 
former case one of the legs needs to be the multiplier field. Interestingly, 
this two-field theory does not allow diagrams beyond one loop. As we shall 
describe below, both (\ref{10}) and (\ref{11}) are related to the target 
space effective action for the open $N{=}2$ string.
\\

\noindent
{\bf 2. N=2 Open Strings\ \ }
The spectrum of world-sheet fields in the NSR formulation of $N{=}2$ strings 
consists of the 2d $N{=}2$ supergravity multiplet, whose conformal gauge fixing
produces the standard set of $N{=}2$ superconformal ghosts, plus $N{=}2$ 
matter fields $(X^{\a\bd},\Psi^{\a\bd})$. The computation of open string 
amplitudes requires the evaluation of correlation functions for 
appropriate choices of physical external states on Riemann surfaces with
handles, boundaries, punctures, and a harmonic $U(1)$ gauge field background
with instantons. The result is to be integrated over the moduli of the
Riemann surface and the $U(1)$ gauge field, and finally one is 
to sum over the topologies labelled by the Euler and $U(1)$ instanton numbers.
The relative cohomology of the BRST operator determines the string external
states, which are annihilated by the commuting $N{=}2$ Virasoro and the
anticommuting $N{=}2$ antighost zero modes.
The resulting spectrum has the following quantum numbers:
\begin{itemize}
\item total ghost number $u\in\bZ$ 
\qquad\qquad $\,\bullet$\ target space momentum $k^{\a\bd}$
\item total picture $\pi\in\bZ$
\qquad\qquad\qquad\quad  $\bullet$\ picture twist $\Delta\in\bR$
\item $gl(1,\bR)\oplus sl(2,\bR)'$\quad quantum numbers\quad $m$, $(j',m')$
\end{itemize}
These quantum numbers are however redundant, since they have interrelationships:
$k{\cdot}k=0$ (i.e. $ k^{\a\bd} {=} \k^\a\k^\bd $ ), $\ u{-}\pi=1$, $\ j'=m'=0$,
and $m$ runs in integral steps from $-j$ to $+j$, where $j:=\fr{\pi}{2}{+}1$.
The physical spectrum consists of just one $SL(2,\bR)'$ singlet 
for each value of $\pi$, $\Delta$, and $(\k,\dot{\k})$.
There is still a certain redundancy, since the pictures $(\pi\ge\pi_0,\Delta)$ 
can be reached from $(\pi_0,0)$ by applying spectral flow~$\cal{S}$ and 
picture raising~$\cp^\a$, which commute with the BRST operator and effect
the mappings
\begin{equation}
(\pi,\Delta)\ \stackrel{\cal{S}(\rho)}{\longrightarrow}\ (\pi,\Delta{+}2\rho)
\qquad,\qquad 
(\pi,\Delta)\ \stackrel{\cp^\a}{\longrightarrow}\ (\pi{+}1,\Delta)
\label{16}
\end{equation}
with $\rho\in\bR$.
Because the string path integral integrates over the twists of the $U(1)$
gauge bundle, it averages over the spectral flow orbits. The  
$\cal{S}$-equivalent states therefore ought to be identified and we may choose
the $\Delta{=}0$ representative.
Picture lowering can also be constructed, except on zero-momentum
states. In essence, all physical states (with $k{\neq}0$) can be generated 
starting from the canonical picture $\pi{=}-2$ (i.e. $j{=}0$), 
and the result is symmetric under the ``Poincar\'e duality'' 
$\pi\stackrel{*}{\rightarrow}-4-2\pi$ (i.e. $j\stackrel{*}{\rightarrow}-j$):
\begin{equation}
\begin{array}{c|ccccccccc}
\pi & \cdots & -5 & -4 & -3 & -2 & -1 & 0 & +1 & \cdots \\
j   & \cdots & -\fr32 & -1 & -\fr12 & 0 & +\fr12 & +1 & +\fr32 & \cdots \\
\hline
{\rm states} & \cdots & \ket{\a\b\g}^* & \ket{\a\b}^* & \ket{\a}^* & 
\ket{\ } & \ket{\a} & \ket{\a\b} & \ket{\a\b\g} & \cdots 
\end{array} \quad.
\label{17}
\end{equation}
The states form $SL(2,\bR)$ tensors of rank $2|j|$ (spin $|j|$),
because the picture-raising operator $\cp^\a$ carries a spinor index.
There is no contradiction with the above statement of unit multiplicity,
since all states in a given $SL(2,\bR)$ multiplet are related to each other, 
albeit in a non-local fashion. The open spinor indices are just
carried by normalisation factors multilinear in $\k^\a$, with $\cp^\a$
increasing the spin by~$\fr12$:
\begin{equation}
\cp^{\a_1}\cp^{\a_2}\ldots\cp^{\a_{2j}} \ket{(0);k}\ =\ 
\ket{\a_1\a_2\ldots\a_{2j} ;k}\ \propto\
\k^{\a_1}\k^{\a_2}\ldots\k^{\a_{2j}}\ \ket{(j);k} \quad.
\end{equation}

The NSR formulation of $N{=}2$ strings introduces a complex structure in the 
target space, which explicitly breaks $SO(2,2)\to GL(1,\bR)\otimes SL(2,\bR)'$.
Individual pieces of an $n$-point amplitude are only $SL(2,\bR)'$ invariant,
and contributions from the $M$-instanton $U(1)$ background carry a $gl(1,\bR)$
weight equal to~$M$. Surprisingly, the path integral measure constrains 
the instanton sum to $|M|\le J\equiv n{-}2$ at tree level. 
Moreover, the weight factors built from the string coupling~$e$ 
and the instanton angle~$\theta$ conspire to restore $SO(2,2)$ invariance 
of the instanton sum
if $\sqrt{e}(\cos\fr{\theta}{2},\sin\fr{\theta}{2})$ is assumed 
to transform as an $SL(2,\bR)$ spinor! This spinor simply parametrises the
choices of complex structure, and it may be Lorentz-rotated to $(1,0)$.
Henceforth we shall remain in such a frame where $e{=}1$ and $\theta{=}0$. 
It has the virtue that only the highest $SL(2,\bR)$ weights 
$m_i{=}j_i$, $i=1,\ldots,n$, occur and 
only the maximal instanton number sector, $M=J$, contributes.

The tree-level open string on-shell amplitudes may then be found to be
\begin{eqnarray}
A_3^{\rm string}\ &=&\ f^{abc}\ \k_1^+\k_2^+\ \k_1^\ad \k_{2\ad}\ 
=\ A_3^{\rm Leznov} 
\label{18} \\
A_4^{\rm string}\ &\propto&\ \k_1^+\k_2^+\k_3^+\k_4^+ 
(\kd_1{\cdot}\kd_2\ \kd_3{\cdot}\kd_4\ t + 
 \kd_2{\cdot}\kd_3\ \kd_4{\cdot}\kd_1\ s)\ =\ 0 
\label{19} \\
A_{n>4}^{\rm string}\ &=&\ 0 \quad .
\label{20}
\end{eqnarray}
They are independent of the external $SL(2,\bR)$ spins $j_i$, as long as 
$\sum_{i=1}^n j_i=J\equiv n{-}2$.
Clearly, the Leznov version (\ref{14}) of SDYM is reproduced. However, a
covariant description needs to take the entire tower of states in (\ref{17})
into account.
\\

\noindent
{\bf 3. Target Space Actions\ \ }
Physical string states correspond to target space (background) fields,
whose on-shell dynamics is determined by the string scattering amplitudes.
In particular, the string three-point functions directly yield cubic terms 
in the effective target space action. In the present case, the correspondence 
reads
\begin{equation}
\ket{++\ldots+}\ \longleftrightarrow\ \f^{--\ldots-} \quad (j{\ge}0)
\quad,\qquad 
\ket{--\ldots-}^*\ \longleftrightarrow\ \f^{++\ldots+}\quad(j{<}0)
\label{21}
\end{equation}
and we denote the fields by $\f_j$.
Then, the target space effective action for the infinite tower $\{\f_j\}$ is
\begin{eqnarray}
S_\infty\ &=&\ \int\!d^4x\; \Tr\biggl\{
-\fr12\sum_{j\in\bZ/2} \f_{(-j)}\bo\f_{(+j)}\ +\ 
\fr13\sum_{j_1+j_2+j_3=1}
\f_{(j_1)}\ \Bigl[ \pa^{+\ad}\f_{(j_2)}\ ,\ \pa^+_{\;\;\ad}\f_{(j_3)}
 \Bigr]\biggr\} \nonumber \\
&=&\ \int\!d^4x\; \Tr\biggl\{ -\fr12\ \F^{--}\bo\F^{--}\ +\ \fr16\ 
\F^{--}\Bigl[ \pa^{+\ad}\F^{--} , \pa^+_{\;\;\ad}\F^{--}\Bigr]
\biggr\}_{\eta^4}
\label{22}
\end{eqnarray}
where we have introduced a ``picture hyperfield'',
\begin{equation}
\F^{--}(x,\eta^-)\ =\ \sum_j\ (\eta^-)^{2j}\ \f_{1-j}(x)\quad,
\label{23}
\end{equation}
depending on an extra commuting coordinate~$\eta^-$, and we project
the Lagrangean onto the part quartic in~$\eta$.
It is remarkable that the action (\ref{22}) has the Leznov form in terms
of the hyperfield. It not only reproduces all (tree-level) string three-point
functions (\ref{18}) but also yields vanishing four- and probably
higher-point functions for the same reason that the Leznov action 
(\ref{10}) does. 
Picture raising induces a dual action on the component fields,
\begin{equation}
Q^+:\quad \f_j\ \longrightarrow\ (3{-}2j)\ \f_{j-\fr12} \quad,
\label{24}
\end{equation}
which is nothing but the $\eta^-$ derivative on the hyperfield!

Three successive truncations to a finite number of fields are possible.
First, keeping only $\{\f_{-1},\f_{-\fr12},\f_0,\f_{+\fr12},\f_{+1}\}$,
a consistent five-field model ensues, viz.,
\begin{equation}
\begin{array}{rl}
S_5\ =\ {\displaystyle{\int}}\!d^4x\; \Tr\biggl\{ &
          \fr12 \pa^{+\da} \f\; \pa^-_{\;\;\da} \f\
                    +\  \pa^{+\da} \f^{+} \pa^-_{\;\;\da} \f^{-}\
                    +\  \pa^{+\da} \f^{++} \pa^-_{\;\;\da} \f^{--}\
          \\[5pt]  &
                 +\ \fr12 \f\; [ \pa^{+\da} \f^{-} , \pa^+_{\;\;\da} \f^{-} ]\
                 +\ \fr12 \f^{--} [ \pa^{+\da} \f\; , \pa^+_{\;\;\da} \f ]
	  \\[5pt]  &
                 +\     \f^{--} [ \pa^{+\da} \f^+ , \pa^+_{\;\;\da} \f^- ]\
                 +\ \fr12\f^{++}[\pa^{+\da}\f^{--} , \pa^+_{\;\;\da}\f^{--}]\
          \biggr\}\quad.  
\label{S5}
\end{array}
\end{equation}
Second, eliminating also the fermions leaves us with three fields.
Third, we may in addition kill $\f_0$ as well, resulting in the 
two-field model of Chalmers and Siegel~\cite{cs}! 
All truncations share the one-loop exactness mentioned before.
\\

\noindent
{\bf 4. Self-Duality in Hyperspace\ \ }
The infinite tower of higher-spin 
fields which arise from the picture degeneracy parametrise simply SDYM
in a hyperspace with coordinates $\{x^{\a\ad},\eta^\a,\bar\eta^\ad\}$,
with $\eta$ and~$\bar\eta$ {\it commuting} spinors. This commutative variant of 
superspace exhibits a $\bZ_2$-graded Lie-algebra variant of the 
super-Poincare algebra (i.e. with all anti-commutators replaced by commutators).
So the covariant target space symmetry is effectively the extension
of the $\bR^{2,2}$ Poincar\'e algebra by two {\it Grassmann-even} spinorial
generators squaring to a translation, i.e., $[ Q_\ad, Q_\a ] = P_{\a\ad}$
(see~\cite{dl} for details).
Hyperspace self-duality allows compact expression in a 
chiral subspace independent of the $\bar\eta$ coordinates.
In terms of chiral subspace gauge-covariant derivatives
$\ \cd_\a = \pa_\a + A_\a(x,\eta)\ $ and 
$\ \cd_{\a\ad} = \pa_{\a\ad} + A_{\a\ad}(x,\eta)\ $,
the self-duality conditions take the simple form
\begin{equation}
[\cd_\a,\cd_\b] = \e_{\a\b} F \quad,\quad
[\cd_\a,\cd_{\b\bd}] = \e_{\a\b} F_\bd \quad,\quad
[\cd_{\a\ad},\cd_{\b\bd}] = \e_{\a\b} F_{\ad\bd} \quad.
\label{29}
\end{equation}
Jacobi identities yield the equations
\begin{equation}
\cd_\a^{\;\;\ad} F_{\ad\bd} =0 \qquad,\qquad
\cd_\a^{\;\;\ad} F_\ad =0 \qquad,\qquad
\cd_{\a\ad} F = \cd_\a F_\ad \quad.
\label{30}
\end{equation}
The first two are respectively the Yang-Mills and Dirac equations for
a SDYM multiplet, and the third implies the scalar field equation 
$\cd^2 F=[F^\ad,F_\ad]$.
All chiral hyperfields have $\eta$ expansions, e.g.
\begin{equation}
A_\a(x,\eta)\ =\ A_\a(x)\ +\ \eta^\b A_{\a\b}(x)\
+\ \eta^\b\eta^\g A_{\a\b\g}(x)\ +\ \ldots \quad.
\label{31}
\end{equation}

Choosing the light-cone gauge, $A^+=0=A^+_{\;\;\ad}$, we note that 
all fields are defined in terms of a generalised Leznov prepotential,
\begin{equation}
A^-\ =\ \pa^+\ \F^{--} \qquad,\qquad 
A^-_{\;\;\ad}\ =\ \pa^+_{\;\;\ad}\ \F^{--} \quad,
\label{32}
\end{equation}
\begin{equation}
F\ =\ \pa^+\pa^+\ \F^{--} \quad,\quad
F_\ad\ =\ \pa^+\pa^+_{\;\;\ad}\ \F^{--} \quad,\quad
F_{\ad\bd}\ =\ \pa^+_{\;\;\ad}\pa^+_{\;\;\bd}\ \F^{--} \quad.
\label{33}
\end{equation}
Since $\pa^-$ does not occur in the above, all fields are determined
by the chiral ($\eta^+$--independent) part of $\F^{--}$. The dynamics is
determined by the remaining constraints
\begin{equation}
[\cd^-_{\;\;\ad},\cd^-_{\;\;\bd}]\ =\ 0 \qquad{\rm and}\qquad
[\cd^-,\cd^-_{\;\;\bd}]\ =\ 0\ ,
\label{34}
\end{equation}
where the former equation is precisely the Leznov equation for $\F^{--}$.
Choosing this to be chiral, $\F^{--}=\F^{--}(x,\eta^-)$, allows identification
with (\ref{23}), with action given by (\ref{22}). The second equation above 
then merely determines the $\eta^-$ dependence of $\F^{--}$.

The restricted system of five fields (\ref{S5}) has the $SO(2,2)$-invariant 
action 
\begin{equation}
S_5^{\rm inv}\ =\ \int\!d^4x\; \Tr\biggl\{
\fr14 g^{\a\b}F_{\a\b}\ +\ \fr13 \chi^\a\cd_{\a\ad}F^\ad\ +\ 
\fr18 \cd^{\a\ad}F\;\cd_{\a\ad}F\ +\ \fr12 F\;[F^\ad,F_\ad] \biggr\}
\label{35}
\end{equation}
where $g^{\a\b}$ and $\chi^\a$ are (propagating) multiplier fields for
$A_{\a\ad}$ and $F_\ad$, respectively. 
The similarity with $N{=}4$ supersymmetric SDYM~\cite{s} is evident, 
however with commuting single-multiplicity fermions replacing multiplicity~$4$ 
anticommuting ones.

To conclude, we note that theories of $N{=}2$ closed as well as $N{=}(2,1)$
heterotic strings are also intimately related to self-dual geometry and 
our covariant hyperspace description generalises to both these cases.

\goodbreak\end{document}